\documentclass[journal, onecolumn]{IEEEtran}
\IEEEoverridecommandlockouts
\usepackage{cite}
\usepackage{amsmath,amssymb,amsfonts}
\usepackage{algorithmic}
\usepackage{graphicx}
\usepackage{textcomp}
\usepackage{xcolor}
\def\BibTeX{{\rm B\kern-.05em{\sc i\kern-.025em b}\kern-.08em
		T\kern-.1667em\lower.7ex\hbox{E}\kern-.125emX}}
	
\begin{document}

\title{Improve Sensing and Communication Performance
	of UAV via Integrated Sensing and Communication\\
}

\author{
	Wangjun Jiang,~\IEEEmembership{Student Member,~IEEE,}
	Ailing Wang,~\IEEEmembership{ Member,~IEEE,} \\
	Zhiqing Wei,~\IEEEmembership{ Member,~IEEE,} 
	Meichen Lai,~\IEEEmembership{Student Member,~IEEE,} \\
	Chengkang Pan,~\IEEEmembership{ Member,~IEEE,} 
	Zhiyong Feng,~\IEEEmembership{Senior Member,~IEEE,}
	Jianjun Liu,~\IEEEmembership{ Member,~IEEE,} 
	\\
	
	\thanks{This project is supported by China Mobile Research Institute - Beijing University of Post and Telecommunications Joint innovation center, Project Number: R207010101125D9.

		Wangjun Jiang, Zhiqing Wei, Meichen Lai and Zhiyong Feng are with Beijing University of Posts and Telecommunications, Beijing, China 100876 (email: \{jiangwangjun, weizhiqing, laimeichen, fengzy\}@bupt.edu.cn).  
		
		Ailing Wang, Chengkang Pan and Jianjun Liu are with China Mobile Research Institute. (email: \{wangailing, panchengkang, liujianjun\}@chinamobile.com)
		
		\emph{Correspondence authors: Zhiqing Wei and Zhiyong Feng.}}
	
}

\maketitle

\begin{abstract}
 
 The unmanned aerial vehicle (UAV) needs to sense the environment to ensure safe flight, and the sensing accuracy and communication delay performance are two important indicators of safe flight. The strategy of using integrated sensing and communication (ISAC) technology to improve the sensing and communication performance is proposed in this paper. On the one hand, the extended kalman filter (EKF) algorithm is adopted to achieve the fusion of communication location information and sensing information to improve the accuracy of target sensing. On the other hand, a Identification Friend or Foe (IFF) method based on ISAC is proposed to reduce communication delay. Compared with the traditional IFF method, the integrated technology used for IFF can realize the radar sensing and communication interrogating functions in parallel, greatly shortening the sensing time. Simulation results show that using ISAC technology, the sensing performance of UAV has been greatly improved, the communication delay can be reduced by up to 50\%, the accuracy of target sensing can be improved by 24.2\% when communication location information and radar sensing information have the same sensing accuracy.

\end{abstract}


%
\IEEEpeerreviewmaketitle

\section{Introduction}
%
%
%
%
  
  The unmanned aerial vehicle (UAV) needs to sense
  the environment to ensure safe flight, and the sensing accuracy and communication delay performance are two important indicators of safe flight.
  
  In terms of sensing, the air target attribute identification is mainly carried out by the method of multi-source information fusion \cite{[Sensor_fusion_1],[Sensor_fusion_2],[Sensor_fusion_3],[Sensor_fusion_4]}. This method can overcome the shortcomings of a single sensor in the identification accuracy, reliability, integrity and other aspects, and can adapt to the complex air combat environment at the same time. Chang {\it{et al.}} proposed a spatial attention fusion method for obstacle detection using mmWave radar and vision sensor \cite{[Chang]}. However, multi-sensing devices will greatly increase the hardware cost and increase the size of the UAV. Communication location information can help improve the sensing accuracy to a certain extent, but the time synchronization of communication location information and sensing information is the premise of fusion. In the traditional UAV system, sensing and communication are realized by different devices respectively, and the information of sensing and communication needs to be synchronized before fusion, which increases the complexity of the system. In this paper, ISAC signals are used to realize radar sensing and communication functions simultaneously, which strictly meets the requirements of time synchronization. Extended kalman filter (EKF) algorithm is used to improve the sensing accuracy to a certain extent \cite{[EKF]}.

  In terms of communication, it should note that the communication delay studied in this paper doesn't represents the communication propagation delay, it includes the entire communication interaction delay when UAV determines whether the target node is a cooperable target. The process can be regarded as a kind of Identification Friend or Foe (IFF) operation, which is to identify and confirm the target's friend or foe attributes through various available technologies and means, combined with general or special sensing and communication platform equipment \cite{[IFF_1]}. Therefore, we use IFF time to denote the communication delay.
  
  IFF is mainly realized through interrogating mechanism, which sends encoded interrogations and receives encoded responses to the transponder in different ways, and these ways have different characteristics \cite{[IFF_code_1]}. 
  In recent years, many domestic and foreign studies have been conducted on the identification of airborne targets as friend or foe \cite{[IFF_3],[Sensor_fusion_4]}. Target detection and attribute identification are required before the identification of friend or foe. At present, the air target attribute identification is mainly carried out by the method of multi-source information fusion \cite{[Sensor_fusion_1],[Sensor_fusion_2],[Sensor_fusion_3],[Sensor_fusion_4]}. This method can overcome the shortcomings of a single sensor in the identification accuracy, reliability, integrity and other aspects, and can adapt to the complex air combat environment at the same time. However, it is designed based on the sensing communication separation system, which requires the serial implementation of target radar detection and communication foe identification process. ISAC implies that the sensing and communication subsystems can realize the sensing and communication functions simultaneously based on ISAC signals \cite{[ISAC_1]}. Therefore, a IFF system based on ISAC can realize target radar detection and communication foe identification process in parallel, thus shortening IFF time.

 \subsection{Main contributions of our work}
 The main contributions of this paper are summarized as follows.

1. Improved IFF method based on ISAC is proposed to reduce communication delay time. Compared with the sensing and communication separation system, the method
based ISAC can realize radar sensing and communication interrogating functions in parallel, greatly shortening the communication delay.

2. The EKF algorithm is adopt to improve the sensing accuracy through communication location information assistance.

3. The improvement of sensing performance based on ISAC is simulated and analyzed. The simulation results show that when the communication location information and radar sensing information have the same sensing accuracy, the target sensing accuracy can be increased by 24.2 \%, and the communication delay can be reduced by 50 \%.

The remaining parts of this paper are organized as follows.
Section \ref{sec:AC_model} describes the anti-collision model of UAV.  Section \ref{sec:Sensing_acc} and section \ref{sec:Com_delay} analyzed the improvement of ISAC technology on sensing accuracy and time, respectively. The improvement of sensing and communication performance based on ISAC is simulated in section \ref{sec:Simulation}. Section \ref{sec:Conclusion} concludes the paper. 

The symbols used in this paper are described as follows. Vectors and matrices are denoted by boldface small and capital letters; the transpose, complex conjugate, Hermitian, inverse, and pseudo-inverse of the matrix ${\bf{A}}$ are denoted by ${{\bf{A}}^{\rm T}}$, ${{\bf{A}}^*}$, ${{\bf{A}}^{\rm H}}$, ${{\bf{A}}^{ - 1}}$ and ${{\bf{A}}^\dag}$, respectively; ${\rm{diag}}(\bf{x})$ is the operation that generates a diagonal matrix with the diagonal elements to be the elements of $\bf x$; $ \otimes $ is the kronecker product operator; $ \odot $ is the hadmard product operator.

Key parameters and abbreviations in this
paper is given in Table \ref{label:abbreviations}.
\begin{table}[ht]
	\caption{Key Parameters and Abbreviations}
	\label{label:abbreviations}
	\begin{tabular}{l|l|l|l}
		\hline \hline
		Abbreviation & Description & Abbreviation & Description  \\ \hline
		IFF &  Identification Friend or Foe & ISAC & Integrated sensing and communication \\ \hline
		UAV & Unmanned aerial vehicle & EKF & Extended kalman filter \\ \hline
		$T_{b,iff}$ & IFF time based on the separation system & $T_{a,iff}$ & IFF time based on ISAC \\ \hline
		$t_{1}$ & The time of sending (radar sensing)/ISAC signals & $t_{2}$ & The time of receiving echoes \\ \hline
		$t_{3}$ & The time of detecting unknown nodes & $t_{4}$ & The time of sending interrogating signals \\ \hline
		$t_{5}$ & The time of decoding interrogation information & $t_{6}$ & The time of receiving decoded information \\ \hline   
		$t_{7}$ & The time of implementing IFF via decoded information & $\rho_t$ & The ratio of shortening communication delay \\ \hline  
		$\rho_s$ & The ratio of improving sensing accuracy \\ \hline  
	\end{tabular}
\end{table}

\section{System Model of UAV}\label{sec:AC_model}

\begin{figure}[ht]
	\includegraphics[scale=0.35]{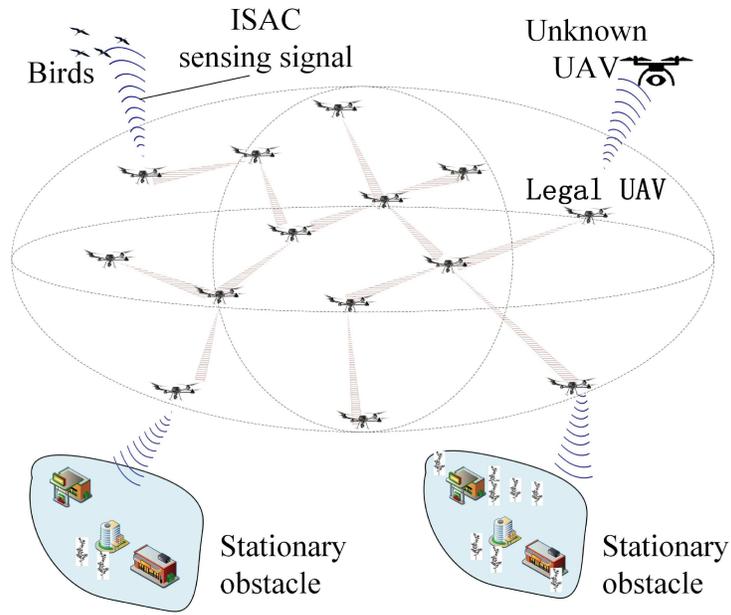}
	\centering
	\caption{UAV swarm sensing environment system.}
	\label{fig:Anti_collision_Model}
\end{figure}

As illustrated in Fig. \ref{fig:Anti_collision_Model}, we assume that there are several UAVs performing tasks in the scene, UAVs sense the environment of the planned area before flying, and a path will be preliminarily generated. However, the UAV may randomly appear dynamic obstacles on the way of flight, and the UAV also needs to avoid all obstacles in the environment and other UAVs. 
Assuming that each UAV in the network is equipped with ISAC capability to detect and identify obstacles in the environment and unnetworked UAVs through ISAC signals.

\section{Shorten the Communication Delay via ISAC}\label{sec:Com_delay}

This section will introduce the scheme of using ISAC technology to shorten the communication delay in the process of target detection. 
The communication delay studied in this paper doesn't represents the communication propagation delay, it represents the entire communication interaction delay when UAV determines whether the target node is a cooperative target. The process can be regarded as a kind of IFF operation. Therefore, we use IFF time $T_{iff}$ to denote the communication delay. In this section, we first compare the IFF time between the sensing communication separation system and ISAC system, then analyze the improvement of the performance of shortening IFF time by adopting ISAC technology. 

\subsection{IFF method Based on Sensing Communication Separation System}\label{sec:Com_delay_1}

\begin{figure}[ht]
	\includegraphics[scale=0.45]{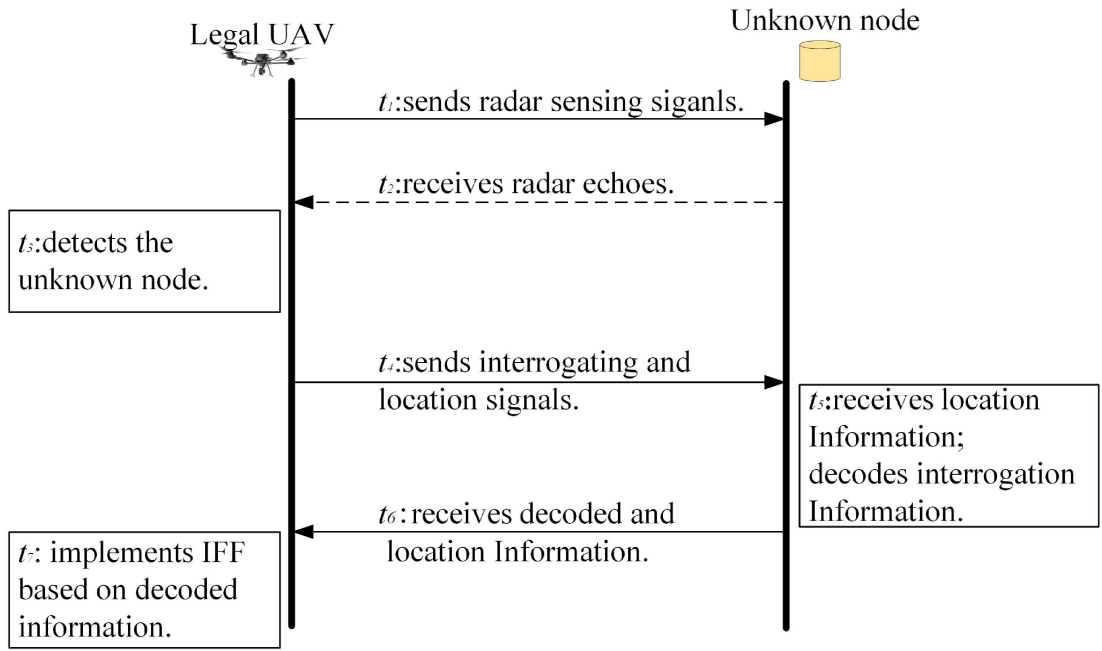}
	\centering
	\caption{IFF time $T_{b,iff}$ based on the sensing and communication separation system.}
	\label{fig:Before_IFF_time}
\end{figure}

In the traditional sensing communication separation system, legal UAVs need to first send radar sensing signals to detect unknown nodes, and then send communication interrogating and location signals to unknown nodes, and complete IFF based on the received response information. As Fig. \ref{fig:Before_IFF_time} shows, the IFF time $T_{b,iff}$ based on the sensing and communication separation system mainly includes the time of sending radar sensing signals $t_{1}$, receiving radar echoes $t_{2}$, detecting the unknown node $t_3$, sending interrogating and location signals $t_4$, decoding interrogation information $t_5$, receiving decoded and location information $t_6$, implementing IFF based on decoded information $t_7$. 

\begin{equation}\label{equ:Before_IFF_time}
T_{b,iff} = t_1 +t_2 + t_3 + t_4 + t_5 + t_6 + t_7.
\end{equation}

\subsection{IFF method Based on ISAC System}\label{sec:Com_delay_2}

\begin{figure}[ht]
	\includegraphics[scale=0.5]{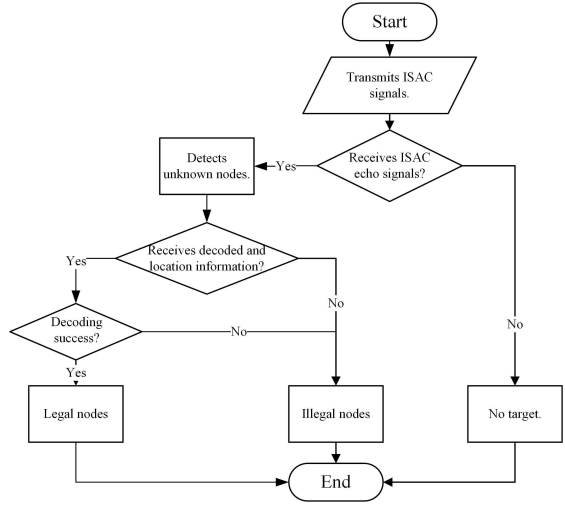}
	\centering
	\caption{IFF method based on ISAC system.}
	\label{fig:After_IFF}
\end{figure}

In the IFF system based on the ISAC system, legal UAVs can transmit ISAC signals to unknown nodes, and receive ISAC echo signals to realize target detection. In the meantime, unknown nodes can obtain interrogation and location information from ISAC signal. The main steps are shown in Fig. \ref{fig:After_IFF}.

\begin{figure}[ht]
	\includegraphics[scale=0.45]{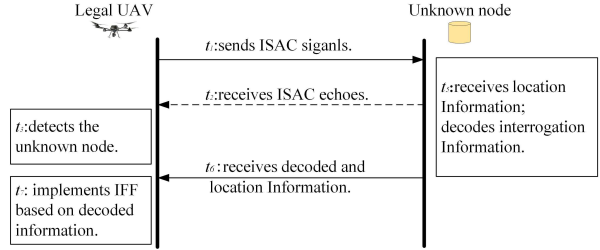}
	\centering
	\caption{IFF time $T_{a,iff}$ based on the ISAC system.}
	\label{fig:After_IFF_time}
\end{figure}

Since the ISAC signal can realize both sensing and communication functions at the same time, the process of decoding interrogation information and detecting the unknown node can be executed in parallel. As Fig. \ref{fig:After_IFF_time} shows,  the IFF time $T_{a,iff}$ based on the ISAC system can be expressed as 

\begin{equation}\label{equ:After_IFF_time}
T_{a,iff} = t_1 + max(t_2 + t_3, t_5) + t_6 + t_7.
\end{equation}

\section{Improve the Sensing Accuracy via Communication Location Information Assistance}\label{sec:Sensing_acc}

This section will introduce a scheme to improve the sensing accuracy by integrating communication positioning information and radar sensing information obtained from ISAC signals.
The sensing error can be reduced to a certain extent by fusion of sensing and communication location information. EKF is a method of least mean square error estimation \cite{[EKF]}, which linearizes the nonlinear system by using first-order Taylor series, which is suitable for radar sensing and communication location information fusion.
According to section \ref{sec:Com_delay}, UAV can obtain target sensing observation vector ${\bf{M}}({\it k})$ and position state vector ${\bf{S}}({\it k})$ through the received echo information and communication positioning information respectively. 
The state vector ${\bf{S}}({\it k})$ contains the position, velocity and acceleration information of the target, which can be expressed as
\begin{equation}\label{equ:EKF_2}
\begin{aligned}
{\bf{S}}({\it k}) = [{\it x}({\it k}), {\it y}({\it k}), {\it v_x}({\it k}), {\it v_y}({\it k}), {\it a_x}({\it k}), {\it a_y}({\it k})]^{\rm{T}}
\end{aligned}.
\end{equation}
The observation vector ${\bf{M}}({\it k})$ contains the distance and direction information of the target, which can be expressed as
\begin{equation}\label{equ:EKF_3}
\begin{aligned}
{\bf{M}}({\it k}) = [{\it R}({\it k}), {\it \alpha}({\it k})]^{\rm{T}}
\end{aligned}.
\end{equation}
Assuming that the position of UAV $A$ at time $k$ is $({\it x}_0({\it k}), {\it y}_0({\it k}))$, then position of the target $({\it x}({\it k}), {\it y}({\it k}))$ can be deduced as 
\begin{equation}\label{equ:EKF_4}
\begin{aligned}
{\it x}({\it k}) = {\it x}_0({\it k}) + {\it R}({\it k}) cos({\it \alpha}({\it k})) \\
{\it y}({\it k}) = {\it y}_0({\it k}) + {\it R}({\it k}) sin({\it \alpha}({\it k}))
\end{aligned}.
\end{equation}
Assuming that the error of ${\bf{M}}({\it k})$ and ${\bf{S}}({\it k})$ follow the gaussian distribution 
\begin{equation}\label{equ:Gaussian_1}
\begin{aligned}
{\bf{M}}({\it k}) &\sim {{\mathcal{CN}}{\rm{(}}0,}{\bf \sigma}_{\bf M}^2 ({\it k}) {\rm{)}} \\
{\bf{S}}({\it k}) &\sim {{\mathcal{CN}}{\rm{(}}0,}{\bf \sigma}_{\bf S}^2 ({\it k}) {\rm{)}} 
\end{aligned},
\end{equation}
where ${\bf \sigma}_{\bf M}^2 ({\it k})$ and ${\bf \sigma}_{\bf S}^2 ({\it k})$ are the covariance matrix of the error.

\begin{figure}[ht]
	\includegraphics[scale=0.5]{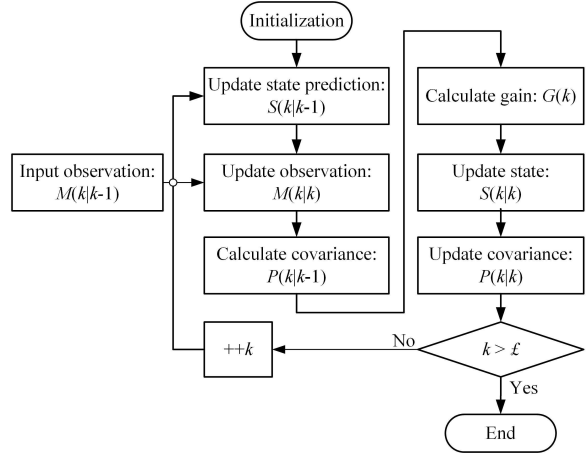}
	\centering
	\caption{EKF algorithm for fusion of sensing and communication location information.}
	\label{fig:EKF}
\end{figure}

As shown in the Fig. \ref{fig:EKF}, the EKF algorithm which used for the fusion of sensing and communication location information mainly includes six steps, detailed derivation can be referred to \cite{[EKF]}.

1) Use the state at time $k-1$ to estimate the state at time $k$:
\begin{equation}\label{equ:EKF_10}
\begin{aligned}
{\bf{S}}({\it k}|{\it k}-1) = \frac{\partial f}{\partial {\bf S}({\it k} | {\it k} - 1)} {\bf{S}}({\it k} -1|{\it k}-1)
\end{aligned},
\end{equation}
where ${\it f}( \cdot )$ is the motion model function of the target.

2) Predict the error covariance matrix:
\begin{equation}\label{equ:EKF_11}
\begin{aligned}
{\bf{P}}({\it k}|{\it k}-1) =& \frac{\partial f}{\partial {\bf S}({\it k} | {\it k} - 1)} {\bf{P}}({\it k} - 1|{\it k}-1) ({\frac{\partial f}{\partial {\bf S}({\it k} | {\it k} - 1)}})^{\rm T} \\
&+ {\bf \sigma}_{\bf S}^2 ({\it k}).
\end{aligned}
\end{equation}

3) Update the observation:
\begin{equation}\label{equ:EKF_12}
\begin{aligned}
{\bf{M}}({\it k}|{\it k}-1) = \frac{\partial h}{\partial {\bf S}({\it k} | {\it k} - 1)} {\bf{S}}({\it k} -1|{\it k}-1)
\end{aligned},
\end{equation}
where ${\it h}( \cdot )$ is the measurement model function of the target.

4) Calculate the filter gain:
\begin{equation}\label{equ:EKF_13}
\begin{aligned}
{\bf{G}}({\it k}) =& {\bf{P}}({\it k}|{\it k}-1) ({\frac{\partial h}{\partial {\bf S}({\it k})}})^{\rm T} [\frac{\partial h}{\partial {\bf S}({\it k})} \cdot \\
& {\bf{P}}({\it k}|{\it k}-1) ({\frac{\partial h}{\partial {\bf S}({\it k})}})^{\rm T} + {\bf \sigma}_{\bf M}^2 ({\it k})]^{-1}.
\end{aligned}
\end{equation}

5) Update state vector:
\begin{equation}\label{equ:EKF_14}
\begin{aligned}
{\bf{S}}({\it k}|{\it k}) = {\bf{S}}({\it k}|{\it k} - 1) + {\bf{G}}({\it k})[{\bf{M}}({\it k}) - {\bf{M}}({\it k}|{\it k}-1)]
\end{aligned}.
\end{equation}

6) Update the error covariance matrix:
\begin{equation}\label{equ:EKF_15}
\begin{aligned}
{\bf{P}}({\it k}|{\it k}) = {\bf{P}}({\it k}|{\it k}-1) -  {\bf{G}}({\it k}) \frac{\partial h}{\partial {\bf S}({\it k})} {\bf{P}}({\it k}|{\it k}-1) 
\end{aligned}.
\end{equation}

\section{Simulation Results}\label{sec:Simulation}

Simulation parameters used in this section are shown in table \ref{Parameter:simulation} \cite{[UAV_parameters_1]}.

\begin{table}[t]
	\caption{Simulation parameters adopted in this paper \cite{[UAV_parameters_1]}.}
	\label{Parameter:simulation}
	\begin{tabular}{l|l|l|l|l|l}
		\hline
		\hline
		Items & Value & Meaning of the parameter & Items & Value & Meaning of the parameter \\ \hline
		$t_3$ & 10 ms & Time of detecting unknown nodes & $t_5$ & (0 $\sim$ 20) ms & Time of decoding interrogation \\ \hline
		$t_7$ & (0 $\sim$ 20) ms & Time of implementing IFF & ${\bf \sigma}_{\bf M}^2$ & $diag(10,0.01)$ & Covariance matrix of ${\bf M}$ \\ \hline
		$v_1$ & 25 $m/s$ & Speed of the UAV & $v_2$ & 25 $m/s$ & Speed of the target \\ \hline
		$a_1$ & 5 $m/s^{\rm 2}$ & Acceleration of the UAV & $a_2$ & 5 $m/s^{\rm 2}$ & Acceleration of the target \\ \hline
		${\it \gamma}_1$ & 150 $^\circ/s$ & Angular speed of the UAV & ${\it \gamma}_2$ & 150 $^\circ/s$ & Angular speed of the target \\ \hline
		$\alpha$ & 5 $^\circ$ & Flight direction of the UAV & $\beta$ & 5 $^\circ$ & Flight direction of the target \\ \hline
		${\bf \sigma}_{\bf S}^2$ & $diag(\sim,\sim,1,1,0.1,0.1)$ & Covariance matrix of ${\bf S}$ \\ \hline
	\end{tabular}
\end{table}

\subsection{Sensing Accuracy}\label{sec:Simulation_2}

\begin{figure}[ht]
	\includegraphics[scale=0.5]{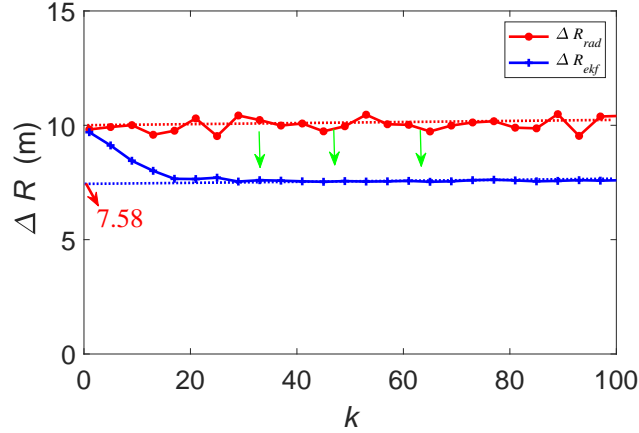}
	\centering
	\caption{$\Delta R_{ekf}$ and $\Delta R_{rad}$ with different iterations $k$.}
	\label{fig:sensing_acc_1}
\end{figure}

\begin{figure}[ht]
	\includegraphics[scale=0.5]{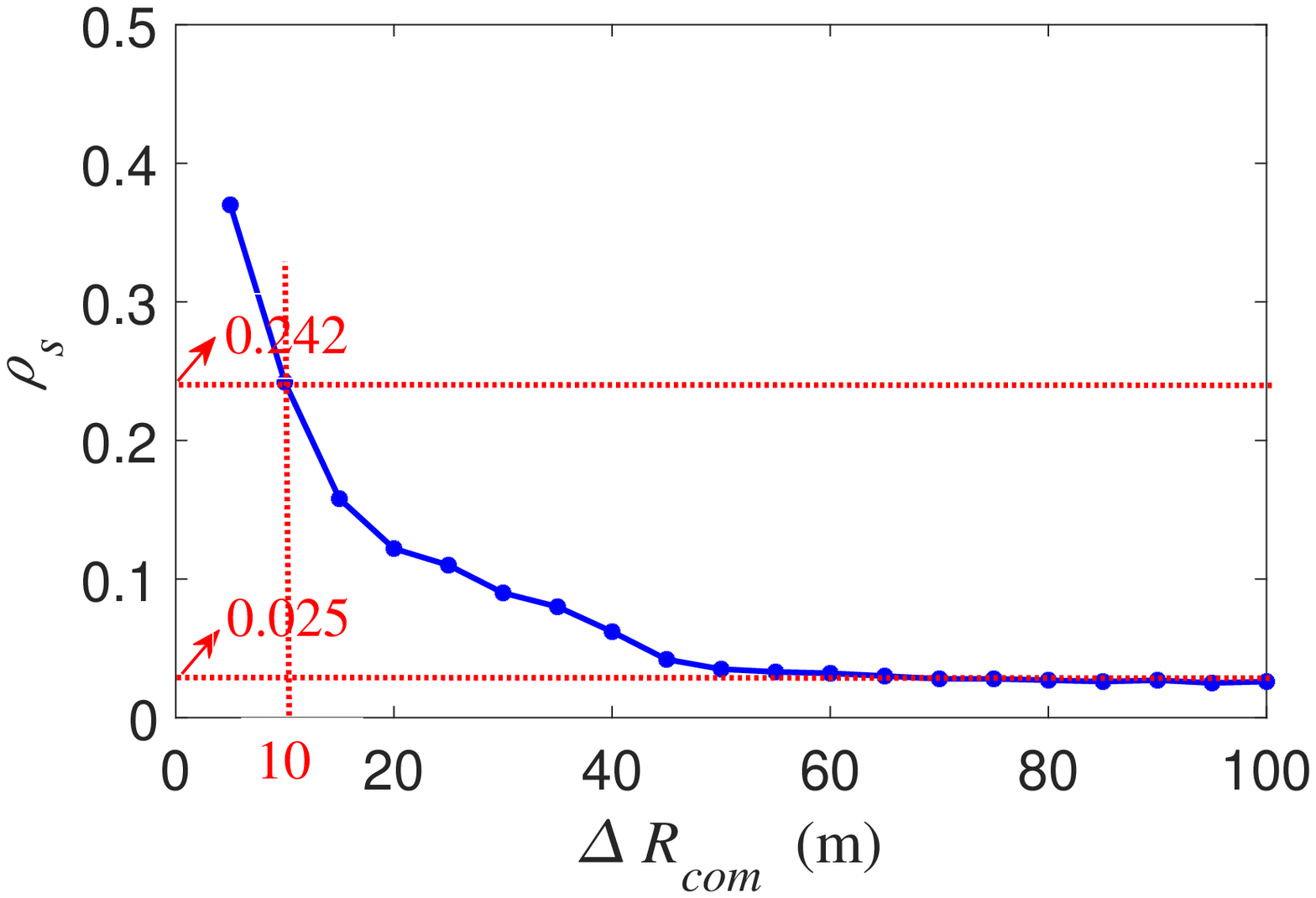}
	\centering
	\caption{$\rho_s$ with different $\Delta R_{com}$ based on $k = 20$.}
	\label{fig:sensing_acc_2}
\end{figure}

According to section \ref{sec:Sensing_acc}, with the help of communication location information, the EKF algorithm can improve the sensing error. Assuming that the error covariance matrix of radar sensing ${\bf \sigma}_{\bf M}^2 ({\it k}) = diag(10, 0.01)$, the error covariance matrix of communication location information ${\bf \sigma}_{\bf S}^2 ({\it k}) = diag(10,10,1,1,0.1,0.1)$. Without loss of generality, the variance of the overall radar sensing error is represented by the range error variance of radar sensing error $\Delta R_{rad}$, the range error variance of communication location information is represented by $\Delta R_{com}$, and the variance of the range error after adopting the EKF algorithm is represented by the variance of the EKF error $\Delta R_{ekf}$. As Fig. \ref{fig:sensing_acc_1} shows, $\Delta R_{ekf}$ decreases with the increase of the number of iterations $k$. When $k \ge 20$, $\Delta R_{ekf}$ almost tends to converge. Moreover, on the whole, $\Delta R_{ekf}$ is smaller than $\Delta R_{rad}$, which verifies the conclusion that the fusion of communication positioning information can assist in improving the sensing accuracy. 

The improvement ratio of sensing accuracy $\rho_s$ can be defined as
\begin{equation}\label{equ:sensing_acc_1}
\begin{aligned}
\rho_s = \frac{\Delta R_{rad} - \Delta R_{ekf}}{\Delta R_{rad}}.
\end{aligned}
\end{equation}
Fig. \ref{fig:sensing_acc_2} shows that when the number of iterations $k = 20$, $\rho_s$ increases with the decrease of $\Delta R_{com}$, and when $\Delta R_{rad} = \Delta R_{com}$, $\rho_s$ can reach 24.2\%.

\subsection{Communication Delay}\label{sec:Simulation_1}

It should be noted that $T_{iff}$ represents only the time taken for one interaction, and multiple interaction are needed to locate the target with higher accuracy. Simulation results in Section \ref{sec:Simulation_2} shows that the improvement of target accuracy tends to be stable after about 20 iterations, so the total communication delay needs to be about $20T_{iff}$. 
Without loss of generality, we can analyze the effect of ISAC technology on communication delay reduction by studying $T_{iff}$.

According to euqation \eqref{equ:Before_IFF_time} and \eqref{equ:After_IFF_time}, compared with IFF based on sensing communication seperated system, the ratio of IFF based on ISAC system shortening IFF time $\rho_t$ can be expressed as

\begin{equation}\label{equ:Shortening_IFF_time_1}
\begin{aligned}
\rho_t &= \frac{T_{b,iff} - T_{a,iff}}{T_{b,iff}} \\
&= \frac{t_2 + t_3 + t_4 + t_5 - max(t_2 + t_3, t_5)}{t_1 +t_2 + t_3 + t_4 + t_5 + t_6 + t_7}.
\end{aligned}
\end{equation}

Compared with the signal processing time, the signal transmission time is very short and can be ignored. Then $\rho_t$ can be simplified as

\begin{equation}\label{equ:Shortening_IFF_time_2}
\begin{aligned}
\rho_t = \frac{t_3 + t_5 - max(t_3, t_5)}{t_3 + t_5 + t_7}.
\end{aligned}
\end{equation}

\begin{figure}[ht]
	\includegraphics[scale=0.5]{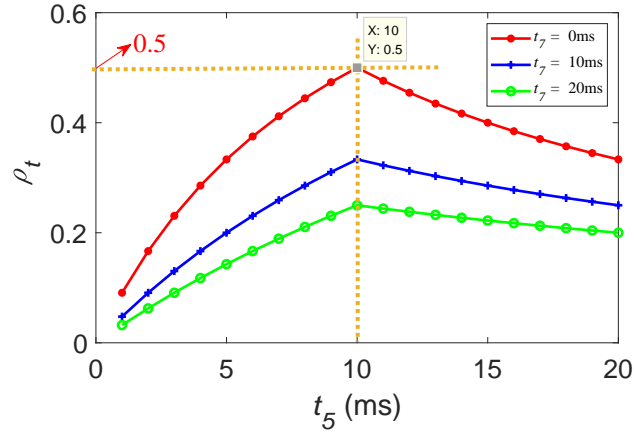}
	\centering
	\caption{$\rho_t$ with $t_3 = 10$ ms.}
	\label{fig:Shortening_IFF_time_1}
\end{figure}

As Fig. \ref{fig:Shortening_IFF_time_1} shows, when the time of detecting the unknown node $t_3$ equals to the time of decoding interrogation information $t_5$, $\rho_t$ can reach the optimal. When the time of implementing IFF based on decoded information $t_7 = 0$ ms, the optimal $\rho_t$ can reach 50\%.

\section{Conclusion}\label{sec:Conclusion}

Target sensing accuracy and communication delay are two important parameters that limit the anti-collision performance, which is important to ensure the flight safety of the UAV. In this paper, we propose a method for reducing IFF time based on ISAC signals. Based on the integration of communication signal perception, target detection and IFF can be executed in parallel, greatly shortens the time of target recognition. Simulation results show that IFF time can shorten up to 50\%. Moreover, we adopt EKF algorithm to improve sensing accuracy with communication location information assistance. Simulation results show that the accuracy of target sensing can be improved by 24.2\% when communication location information and radar sensing information have the same error, which greatly improve the flight safety of UAV.


%

\bibliographystyle{IEEEtran} 
\bibliography{reference}	


\ifCLASSOPTIONcaptionsoff
  \newpage
\fi

\end{document}